\documentclass{aa}    % LaTeX A&A  Standard Fonts
\def\ut#1{\mathop{\vtop{\ialign{##\crcr
     $\hfil\displaystyle{#1}\hfil$\crcr\noalign
     {\kern1pt\nointerlineskip}\hbox{$\hfil\sim\hfil$}\crcr
     \noalign{\kern1pt}}}}}

\def\undersymbol#1#2{\mathop{\vtop{\ialign{##\crcr
     $\hfil\displaystyle{#2}\hfil$\crcr\noalign
     {\kern1pt\nointerlineskip}\hbox{$\hfil#1\hfil$}\crcr
     \noalign{\kern1pt}}}}}

\begin{document}
\thesaurus{02.07.1; 02.07.2; 12.07.1}
\title{A note on gravitational wave lensing}
\author{F. De Paolis$^1$, G. Ingrosso$^1$, A.A. Nucita$^1$ and Asghar Qadir$^{2,3}$}
\institute{Dipartimento di Fisica, Universit\`a di Lecce, and
INFN, Sezione di Lecce,
Via Arnesano, CP 193, 73100 Lecce, Italy\\
2 Department of Mathematical Sciences, King Fahd University of
Petroleum and
Minerals, Dhahran 31261, Saudi Arabia\\
3 Department of Mathematics, Quaid-i-Azam University, Islamabad,
Pakistan} \offprints{F. De Paolis}
\date{Received date; accepted date}
\authorrunning{De Paolis et al.}
\titlerunning{A note on gravitational wave lensing}
\maketitle
\begin{abstract}

In a recent paper (Ruffa \cite{ruffa}) it was proposed that the
massive black hole at the Galactic center may act as a
gravitational lens focusing gravitational wave energy to the
Earth. Considering the gravitational wave signal emitted by
galactic spinning pulsars, an enhancement in the gravitational
wave intensity by a factor of a few thousand is found. For
galactic and extra-galactic sources the intensity enhancement can
be as high as 4,000 and 17,000, respectively. In this note we
consider the probability of significant signal enhancement from
galactic and extra-galactic pulsars by the proposed mechanism and
find that it is actually negligible. The lensing effect due to a
possible companion object (a star or the galactic center black
hole) of the gravitational wave source is also investigated in
the framework of the classical microlensing theory.
\end{abstract}

\section{Introduction}

Gravitational lensing of electromagnetic waves is a well known
phenomenon predicted by the General Theory of Relativity  (for a
review on this issue see Schneider, Ehlers and Falco \cite{sef}).
In principle, gravitational lensing of gravitational waves should
occur in the same way as it does for light. The most obvious
difference is that gravitational wave propagation is not
disturbed by dust grains, as happens for light, so that the
central part of our galaxy may be investigated by using the next
generation of gravitational wave detectors.

In a very interesting paper Ruffa (\cite{ruffa}), assuming that the mass of
the Galactic center is in the form of a massive black hole with mass $%
M\simeq 2.6\times 10^6$ M$_{\odot}$ (for a Galactic center
overview see Eckart, Genzel, Ott and Schoedel \cite{egos}), the
gravitational wave lensing problem was studied by a typical
Fraunhoffer diffraction approach. It was pointed out (Ruffa
\cite{ruffa}) that extra galactic sources can be
amplified by a factor of about $17,000$ and galactic neutron stars by over $%
4,000$. The author also argued that the Earth would take about
10.1 days to traverse the focused region of the extra galactic
sources. For galactic bulge sources the focused region would be
scaled down by a factor of nearly 3 and hence the observing time
would be reduced to about 3 days.

The question that naturally arises is ``how likely is it that we see the
proposed enhancement?'' To answer this question we need to estimate the
number of sources that could be expected to be observed, both galactic and
extra-galactic. In principle, of course, there are many extra-galactic
sources as there are $\sim 10^{11}$ galaxies. However, we need to limit the
number, applying a cut-off by requiring that the expected intensity (after
amplification) be greater than the sensitivity of the detector.

As will be seen, the chances of seeing the dramatic enhancement calculated
by Ruffa are extremely small. This is due to the fact that Ruffa assumes a
very special geometry, with the source, lens and observation point aligned.
Whereas Ruffa considered Fraunhoffer diffraction, an analysis using
``geometric optics'' for gravitational waves had been undertaken, in which
the special geometry of Ruffa was not assumed (De Paolis, Ingrosso and
Nucita \cite{din}). This analysis gave a much higher probability but a lower
enhancement.

In this paper, we evaluate the probability of enhancement of the
gravitational wave signals as a consequence of diffraction by massive
compact objects in the highly aligned geometry.

We also consider the lensing effect due to a possible companion star of the
gravitational wave source in the framework of the classical microlensing
theory. In addition to this case we consider the gravitational wave lensing
due to the massive black hole at the galactic center. As we shall see, the
gravitational wave signal amplification due to the companion object (or the
black hole at the galactic center) might be detectable by the VIRGO detector
for some orbital parameters of the binary system.

\section{Probability of signal enhancement by the galactic central black hole}
Let us assume that the observer, the lens and the source (a
spinning pulsar) are on the same line. The lens will focus the
gravitational waves that pass through the Einstein radius
$R_E=\sqrt{\frac{4GM}{c^2}\frac{Dd}{D+d}}$ to the Earth in a
region orthogonal to the line of sight. Here $D$ is the distance
between the observer and the lens ($\simeq 8$ kpc) and $d$ is the
lens-source distance. Using diffraction theory, the focused area
is determined by $D$ and the diffraction pattern for the circular
annular region bounded by radii $R_E$ and $R_E$+$\delta R_E$.
Note that a gravitational wave that passes at the distance
$R_E$+$\delta R_E$ from the lens is focused in $D+\delta D$. It
is possible to show (by evaluating the 3 dB points of the signal,
see Ruffa \cite{ruffa}) that the radius of the focused region is
given by
\begin{equation}
\rho(d) \simeq 0.175\lambda \frac{D}{R_E(d)}  \label{rhod}
\end{equation}
where $\lambda$ is the gravitational wave length \footnote{In the
case of non-precessing pulsars rotating with frequency of $\sim
30$ Hz, the emitted gravitational waves has frequency $\sim 60$
Hz (see e.g. Shapiro and Teukolsky \cite{st}). Consequently, the
gravitational wave length turns out to be $\sim5\times 10^{8}$
cm.}. As pointed out by Ruffa, the increase in intensity depends
on the ratio (known as the intensity ratio) of the annular region
bounded by the radii $R_E$ and $R_E$+$\delta R_E$ to the focused
region area. This can be determined by the system of equations
\footnote{We note that in all case of interest $R_E\sim 10^{17}$
cm and, from eq. (\ref{rhod}), $\rho\ll R_E$.}
\begin{equation}
\left\{\begin{array}{l}
\rho(d)(D+\delta D)\simeq R_E(d) \delta D~, \\ \\
\delta R_E(d)=\displaystyle{\frac{1}{R_E(d)}\frac{2GM}{c^2}\frac{d^2}{(D+d)^2%
}\delta D}~.\label{system}
\end{array}\right.
\end{equation}
We note that the first equation is Ruffa's eq. (4) at first order
approximation (since $\rho\ll R_E$) and the second one derives
from differentiating the expression for the Einstein radius $R_E$
with respect to $D$. Solving for $\delta D$, we finally obtain
\begin{equation}
\delta R_E(D)\simeq \frac{1}{2}\frac{\rho d}{D+d}~,  \label{deltaRE}
\end{equation}
which reduces to $\delta R_E(D)\simeq \rho/2$ for $d\gg D$.
Taking into account the previous equations and the expression of
the Einstein radius, the intensity ratio turns out to be
\begin{equation}
IR(d)\simeq \frac{d}{D+d}\frac{R_E}{\rho}~,  \label{ir}
\end{equation}
implying a gravitational wave amplification factor \footnote{
Since the gravitational wave amplitude $h$ is proportional to the
square root of the energy flux, the enhancement in $h$ is
proportional to the square root of the intensity ratio.}
\begin{equation}
A=\sqrt{IR(d)}~.  \label{amplification}
\end{equation}
\begin{table*}[tbp]
\caption{The maximum flux amplification $A_{f, max}$, the
gravitational wave
signal amplification $A_{h, max}$ and the microlensing event time-scale $%
T_{1/2}$ are shown for different values of the orbital radius
$R_{orb}$, period $P$ and inclination angle $i$, and the
companion mass object $M$. Here we assume $m=1.4 ~M_{\odot}$. In
the last three lines the black hole at the galactic center is
assumed to be the lens object around which a gravitational wave
emitting pulsar rotates.}
\label{tabella}%\centering
%\medskip
\par
\begin{center}
\begin{tabular}{|l|l|l|l|lll|}
\hline M $(M_{\odot})$ & $R_{orb}$ (cm) & $P (h)$ & $i^o$ &
$A_{f, max}$ & $A_{h, max}$ & $T_{1/2}$ (h) \\ \hline
1.4 & $2\times 10^{11}$ & 8 & 0.1 & 1.46 & 1.21 & $4.4\times 10^{-3}$ \\
1.4 & $2\times 10^{11}$ & 8 & 0.001 & 116.73 & 10.80 & $1.1\times
10^{-4}$
\\
1.4 & $2\times 10^{13}$ & $8\times 10^3$ & 0.1 & 1.0004 & 1.0002 & 3 \\
1.4 & $2\times 10^{13}$ & $8\times 10^3$ & 0.001 & 11.7 & 3.4 &
$8\times 10^{-2}$ \\ \hline
10 & $2\times 10^{13}$ & $4\times 10^3$ & 0.1 & 1.013 & 1.006 & 1.55 \\
10 & $2\times 10^{13}$ & $4\times 10^3$ & 0.001 & 31.20 & 5.58 &
$4\times
10^{-2}$ \\
10 & $2\times 10^{14}$ & $1.2\times 10^5$ & 0.1 & 1.00018 & 1.00009 & 45 \\
10 & $2\times 10^{14}$ & $1.2\times 10^5$ & 0.001 & 9.9 & 3.14 & 1.31 \\
\hline $2.6\times10^6$ & $3\times 10^{16}$ & $5.1\times 10^5$ & 1
& 1.03 & 1.02 &
2047 \\
$2.6\times10^6$ & $3\times 10^{16}$ & $5.1\times 10^5$ & 0.5 &
1.21 & 1.10 &
1210 \\
$2.6\times10^6$ & $3\times 10^{16}$ & $5.1\times 10^5$ & 0.1 &
4.14 & 2.04 & 378 \\ \hline
\end{tabular}
\end{center}
\end{table*}
Assume now that Ruffa's treatment still holds if the source is
not fully aligned with the observer-lens line but forms at least
an angle $\theta \simeq \rho(d)/d$ with it. Consequently, the
probability density of finding the source within a circular
region of radius between $\rho$ and $\rho + d\rho$ is
proportional to the ratio between that circular area and the
surface of the cylinder centered at the galactic center and
having radius $d$ and height $H$. From simple geometric
considerations, this probability density, after integration over
$d\rho$, gives the following probability of finding a source
within the radius $\rho(d)$
\begin{equation}
P_g(d)\simeq 3\times 10^{-3} \frac{c^2}{GM}\lambda ^2 \frac{D(D+d)}{d^2H}~.
\label{probdistanza}
\end{equation}

On the other hand, the real probability to observe a gravitational wave
enhancement in the case of the observer, lens and source on the same line of
sight has to take into account also the number of expected sources behind
the central black hole.

From an extrapolation of the neutron star (NS) birthrate (Narayan
and Ostriker \cite{no}) \footnote{From historical supernovae
(SNe) in our galaxy one gets an SN frequency of one event every
$40\pm 10$ yr (Tammann, Loeffler and Schroeder \cite{tls}). These
estimates do not take into account any SN rate evolution with
time and the fact that only about $10\%$ of the historical SNe
have been close and bright enough to have been detectable  by the
naked eye, and therefore have to be considered as a lower limit to
the inferred present galactic SN number. A more realistic
estimate of the SN rate for our galaxy is about one over 20 years
(Panagia \cite{panagia}). Observations show that large late-type
spiral galaxies produce about 10 SNe per century (Tammann,
Loeffler and Schroeder \cite{tls}).} and from the number of
supernovae required to account for the heavy element abundance in
the Milky Way (Arnett, Schramm and Truran \cite{ast}), we expect
that our galaxy contain $N\simeq 10^9$ NSs. For simplicity we
assume that the NS population has a usual disk-like shape, i.e.
the number density profile is given by
\begin{equation}
n(r,z)\simeq\frac{\Sigma_{NS}}{2Hm}e^{-{z}/{H}}
                                   e^{-{(r-D)}/{\Delta}}~,  \label{NSnumerdensity}
\end{equation}
where $\Sigma_{NS}\simeq 1$ M$_{\odot}$ pc$^{-2}$, $m\simeq 1$
M$_{\odot}$, $H\simeq 0.30 $ kpc and $\Delta \simeq 3.5$ kpc is
the typical length scale of the galactic disk. In order to
consider the distribution of matter behind the black hole,
equation (\ref{probdistanza}) has to be multiplied by the factor
$n(r,0) V/N$, $V$ being the total volume of the disk.
Consequently, the probability to have an enhancement of the
gravitational wave signal according to the mechanism proposed by
Ruffa is
\begin{equation}
P(d)\simeq 10^{-2}\frac{\Sigma_{NS}}{Nm} \frac{c^2}{GM}\frac{
\lambda^2\Delta^2}{H}\frac{D(D+d)}{d^2}e^{-(d-D)/\Delta}~{\rm
yr}^{-1}.
\end{equation}

\begin{figure}[tbph]
\begin{center}
\vspace{6.4 cm} \includegraphics{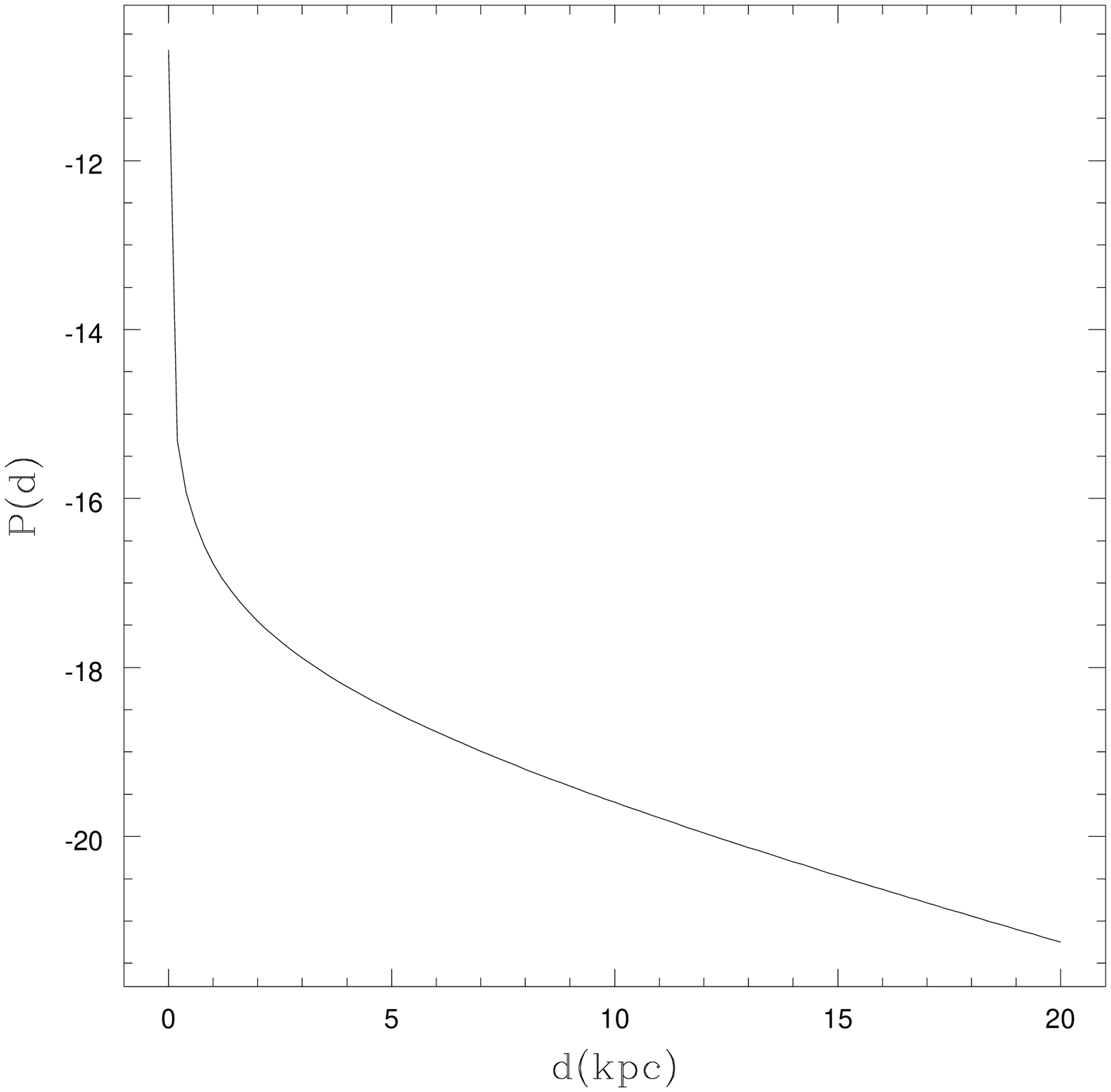}
\end{center}
\caption{The probability of observing the gravitational wave signal
amplitude enhancement by the black hole at the galactic center from a
galactic disk pulsar is shown as a function of the source distance from the
lens.}
\label{fig1}
\end{figure}

In Fig. \ref{fig1} we give the probability $P$ as a function of the source
distance $d$ from the galactic center.
%%%%%%%%%%%%%%%%%%%%%%%%%%%%%%%%%%%%%%%%
%%%%%%%%%%%%%%%%%%%%%%%%%%%%%%%%%%%%%%%%
As one can see, the probability to have a gravitational wave
enhancement \textit{a la Ruffa} by the black hole at the galactic
center from a pulsar in the line of sight to the galactic center
is as low as $\sim 10^{-20}$ per year. The total probability
(assuming that there are no pulsars closer than $5\times 10^{-5}$
kpc to the galactic center) is $10^{-12}$ per year. In other
words, we could expect to wait for a hundred times the age of the
universe to see one!

Let us see now what happens in the case of a gravitational wave
signal due to a supernova explosion in other galaxies behind the
galactic black hole. Since $d\gg D$, in this case $\rho \simeq
0.175 \lambda \sqrt{Dc^2/4GM}$. The probability of a galaxy being
within the focused region at a distance $d$ is simply
$(\rho/2d)^2$. Thus the total probability will go as
$(\rho/d_{min}-\rho/d_{max})/4$. Since $d_{max}>d_{min}$, we only
need to find the probability of a galaxy at the closest distance
to be on the line of sight, i.e. the number density of galaxy
does not matter. It is known that supernovae explode in typical
galaxies less frequently than once every $25$ years.
Consequently, the relevant probability is $\sim 0.01\rho/d_{min}$
per year. Taking $d_{min}$ as 1 Mpc, the probability is less than
$\sim 10^{-15}$ per year! As such, there is no real likelihood of
being able to see the Ruffa enhancement. Of course, there are
also other expected sources of bursts of gravitational waves from
gamma-ray bursts (see e.g. Ruffert and Janka \cite{rj})  or binary
systems (see e.g. De Paolis, Ingrosso and Nucita \cite{din2} and
references therein) that could in principle be detectable by
VIRGO (Jaranowski and Kr\`olak \cite{jk} and Passaquieti
\cite{passaquieti}), LIGO (Barish \cite{barish}) or LISA
(Hiscock, Larson, Routzahn and Kulick \cite{hlrk} and Vecchio
\cite{vecchio}) experiments. However, from the discussion above,
it can easily be calculated that the expected probability of
seeing a gravitational wave amplification by the central black
hole of one of such event is even less than $\sim 10^{-15}$ per
year.

\section{Signal enhancement in binary systems}

Let us consider now the enhancement of the gravitational waves emitted by a
rotating neutron star due to the microlensing effect of a companion star
moving around the compact source on a circular orbit with radius $R_{orb}$.
Let $m$ and $M$ be the masses of the pulsar and of the companion star,
respectively. We further assume that the orbital plane of the binary star is
inclined at an angle $i$ with respect to the line of sight. The companion
star coordinates, with respect to a reference frame with $x$ and $y$ axes on
the orbital plane, are given by
\begin{equation}
x_{M}=R_{orb}\cos \theta ~,~~~~~y_{M}=R_{orb}\sin \theta ~,~~~~~z_{M}=0~,
\end{equation}
where $\theta =\sqrt{G(m+M)/R_{orb}^{3}}t$ is the orbital angle measured
with respect to the positive direction of the $x$ axis pointing towards the
observer. The distance between the companion star $M$ and the line of sight
to the gravitational wave source turns out to be
\begin{equation}
d_{ls}(\theta )=\sqrt{x_{M}^{2}\sin ^{2}i+y_{M}^{2}}.  \label{dist}
\end{equation}
For a source-observer distance greater than the distance between
the source and the lens ($\sim R_{orb}$), the Einstein radius is
$R_{E}\simeq \sqrt{4GMR_{orb}/c^{2}}$. Consequently, as in usual
microlensing theory, the flux amplification factor is given by
(see e.g. Schneider, Ehlers and Falco \cite{sef})
\begin{equation}
A_{f}=\frac{u^{2}+2}{u\sqrt{u^{2}+4}}~,
\end{equation}
with $u=d_{ls}(\theta )/R_{E}$. Obviously, since the impact
parameter $u$ depends on the position of the companion star on
the orbit, we expect that the flux amplification $A_{f}$ is a
periodic function of time with period $P=2\pi
\sqrt{\frac{R_{orb}^{3}}{G(m+M)}}$. We also note that the effect
of the amplification is maximal for $i\rightarrow 0^{o}$. The
expected gravitational wave amplitude enhancement is $A_{h}\simeq
\sqrt{A_{f}}$. In Table \ref{tabella}, assuming $m=1.4~M_{\odot
}$, the maximum flux amplification $A_{f,max}$, the gravitational
wave signal amplification $A_{h,max}$ and the microlensing event
time-scale $T_{1/2}$ are shown for different values of the
orbital parameters ($R_{orb}$, $P$, $i$ and $M$). As one can
note, the amplification effect decreases for increasing
inclination angles while the event time-scale $T_{1/2}$ goes up.
Since the gravitational wave amplification is the same for a
compact or a normal companion star, assuming that VIRGO will be
able to detect only $1\%$ of the pulsars with spin period
$P_{sp}<100$ ms within about $25$ kpc from Earth in three years
of integration, one expects $\sim 10^{4}$ detectable binaries.
Therefore, the number of binary systems with plane inclination
angle $i\leq 0.1^{0}$ is easily found to be about 10. Of course,
since the probability of having high enough amplification of the
signal for long enough is expected to be rather small, the
chances of having a really detectable amplified signal from a
binary system are low. Some higher probability is expected if the
lens object is the black hole at the galactic center, due to its
higher mass. In this case, as one can see from Table
\ref{tabella}, both the signal amplification and duration are
expected to be high enough to have a signal detectable by VIRGO.
For details about the detectable gravitational wave amplitude see
Gourgoulhon and Bonazzola \cite{gb}.

Finally, we want to note that the effects of the gravitational
light deflection and amplification in eclipsing binary stars has
been considered in detail (Marsh \cite{marsh}, Beskin and Tuntsov
\cite{bt}) and by Schneider (\cite{schneider}) for the binary
pulsar 1957+20 as a way for determining the binary system
parameters. Actually, the gravitational wave amplification is, if
gravitational wave astronomy becomes practicable in the future, a
much more powerful tool since gravitational waves are never
shielded by the lens object (as happens for light rays) giving
the unique possibility to trace the sources even in regions (like
the galactic center) where light suffers strong absorption.

\acknowledgements{One of us (AAN) is grateful to KFUPM for
hospitality and cordiality. AQ would also like to acknowledge the
research facilities of KFUPM and the Physics Department of the
University of Lecce, where this work was initiated.}


\begin{thebibliography}{1989}
\bibitem[1989]{ast}  Arnett W.D., Schramm D.N. and Truran J.W., 1989, ApJ,
339, L25
\bibitem[2000]{barish}
Barish B.C., 2000, Advances in Space Research, Volume 25, 1165
\bibitem[2002]{bt} Beskin G.M. and Tuntsov A.V., 2002, preprint
astro-ph/0208095
\bibitem[2001]{din}  De Paolis F., Ingrosso G. and Nucita A.A., 2001, A\&A 366,
1065
\bibitem[2002]{din2} De Paolis F., Ingrosso G. and Nucita A.A., 2002,
A\&A 388, 470
\bibitem[2002]{egos}  Eckart A., Genzel R., Ott T. and Schoedel R., 2002,
preprint astro-ph/0201031
\bibitem[1996]{gb}  Gourgoulhon E. and
Bonazzola S., 1996, Proceedings of the International Conference
on Gravitational Waves: Sources and Detectors, Italy: World
Scientific
\bibitem[2000]{hlrk}
Hiscock W.A., Larson S.L., Routzahn J.R. and Kulick B., 2000, ApJ
540, L5
\bibitem[1999]{jk} Jaranowski P. and Kr\`olak A., 1999,
Phys.Rev. D59, 063003
\bibitem[2001]{marsh} Marsh T. R., 2001, MNRAS 324, 547
\bibitem[1990]{no}  Narayan R. and Ostriker J.O.M., 1990, ApJ, 352, 222
\bibitem[1999]{panagia} Panagia N., 1999, Proceedings of the
International Summer School on {\it Experimental Physics of
Gravitational Waves} (Eds. Barone M. et al.), World Scientific,
pag. 107
\bibitem[2000]{passaquieti} Passaquieti R., 2000, Nuclear Physics B 85,  241
\bibitem[1999]{ruffa} Ruffa A.A. ApJ 517, L31, 1999
\bibitem[1998]{rj} Ruffert M. and Janka H.T., 1998, A\&A 338, 535
\bibitem[1989]{schneider} Schneider J., 1989, A\&A 214, 1
\bibitem[1992]{sef}
Schneider P., Ehlers J. and Falco E.E., 1992, Gravitational
Lenses, Cambridge: Springer-Verlag
\bibitem[1983]{st}
Shapiro S.L. and  Teukolsky S.A., 1983, {\it Black Holes, White
Dwarfs, and Neutron Stars}, John Wiley and \& Sons, New York
\bibitem[1994]{tls} Tammann G. A., Loeffler W. and Schroeder A.,
1994, ApJS 92, 487
\bibitem[1999]{vecchio}
Vecchio A., 1999, in: {\it Gravitational Waves, Third E. Amaldi
Conference}, AIP Conference Proceedings, vol. 523 p. 238, AIP
Press
\end{thebibliography}
\end{document}